\documentclass[12pt]{iopart}

\usepackage{iopams}
\expandafter\let\csname equation*\endcsname\relax
\expandafter\let\csname endequation*\endcsname\relax
\usepackage{graphicx}
\usepackage{color}
\usepackage{amssymb,amsfonts,amsmath}
\usepackage{epstopdf}
\usepackage{subfigure}
\usepackage{braket}
\usepackage{lipsum}
\usepackage{bm}
\usepackage[normalem]{ulem}
\definecolor{orange}{rgb}{1,0.5,0}

\newcommand{\Atan}{Atan}
\newcommand{\Mod}{mod}

\renewcommand{\footnoterule}{%
  \kern -3pt
  \hrule width \columnwidth height 1pt
  \kern 2pt
}

\begin{document}

\title{Desynchronization and pattern formation in a noisy feedforward oscillators network}

\author{Cl\'{e}ment Zankoc} 
\address{Dipartimento di Fisica e Astronomia and CSDC, Universit\`{a} degli Studi di Firenze, via G. Sansone 1, 50019 Sesto Fiorentino, Italia}
\address{INFN Sezione di Firenze, via G. Sansone 1, 50019 Sesto Fiorentino, Italia}
\address{Department of Physics and Institute for Complex Systems and Mathematical Biology, Kings College, University of Aberdeen, Aberdeen AB24 3UE, United Kingdom}
\author{Duccio Fanelli}  \address{Dipartimento di Fisica e Astronomia and CSDC, Universit\`{a} degli Studi di Firenze, via G. Sansone 1, 50019 Sesto Fiorentino, Italia}
\address{INFN Sezione di Firenze, via G. Sansone 1, 50019 Sesto Fiorentino, Italia}
\author{Francesco Ginelli} 
\address{Department of Physics and Institute for Complex Systems and Mathematical Biology, Kings College, University of Aberdeen, Aberdeen AB24 3UE, United Kingdom}
\author{Roberto Livi} 
\address{Dipartimento di Fisica e Astronomia and CSDC, Universit\`{a} degli Studi di Firenze, via G. Sansone 1, 50019 Sesto Fiorentino, Italia}
\address{INFN Sezione di Firenze, via G. Sansone 1, 50019 Sesto Fiorentino, Italia}

\begin{abstract}
We consider a one-dimensional directional array of diffusively coupled oscillators. They are perturbed by the injection of a small additive noise, typically orders of magnitude smaller than the oscillation amplitude, and the system is studied in a region of the parameters that would yield  deterministic synchronization. Non normal directed couplings seed a coherent amplification of the perturbation:  this latter manifests as a modulation, transversal to the limit cycle, which gains in potency node after node.  If the lattice extends long enough, the initial synchrony gets eventually lost and the system moves toward a non trivial attractor, which can be analytically characterized as an asymptotic splay state. The noise assisted instability, ultimately vehiculated and amplified by the non normal nature of the imposed couplings, eventually destabilizes also this second attractor. This phenomenon yields spatiotemporal patterns, which cannot be anticipated by a conventional linear stability analysis.
\end{abstract}

\maketitle

\section{Introduction}

Understanding the origin and functional significance of self-organized patterns of activity, is a challenging question of broad applied and fundamental importance \cite{kuramoto,nicolis}. In many realms of investigation, the system under inspection is composed by individual excitable units, which execute periodic oscillations \cite{goldbeter}. Often, coupling together an ensemble made of identical oscillators can eventually yield a fully synchronized solution \cite{pikovsky}. This amounts to operate the system in unison, the oscillations displayed on different sites of the collection being perfectly coordinated, with no phase delay.  For many applications of interest, as e.g. the study of collective oscillations in neuroscience, distinct deterministic oscillators occupy the nodes of a heterogenous network, which defines the embedding structural support \cite{latora,barrat}. Diffusive couplings between adjacent mesoscopic units are customarily assumed, a paradigmatic choice which proves adequate in many cases \cite{cross}, from modeling the electrical synapses to problems related to the energy management in power plants. Instabilities, however, may be triggered by the punctual injection of a heterogeneous perturbation \cite{nakao}, a tiny source of stochastic disturbance which, under specific conditions, amplifies and eventually breaks the oscillators' synchrony \cite{pecora}. The instabilities instigated by random fluctuations are often patterns precursors \cite{koseska, fanelli_burioni}. The imposed perturbation materializes in fact in patchy motifs of the concentration amount, characterized by a vast gallery of shapes and geometries. An archetypal model of self-sustained oscillations is the celebrated Complex Ginzburg-Landau equation (CGLE),
often evoked as a pillar of non linear phenomena,  from superconductivity to superfluidity and Bose-Einstein condensation, via strings in field theory and neuroscience \cite{aranson}. The CGLE, defined on ordinary or graph-like supports, admits a time-dependent uniform synchronized solution, of the limit cycle type. Deviations from a periodic waveform, sustained by nonlinearities, yield a prototypical modulational instability characterized by  spectral-sidebands and the breakup of the waveform into a train of pulses. This is the so called Benjamin-Feir (BF) instability, named after the researchers who first identified the phenomenon working with periodic surface gravity waves (Stokes waves) on deep water \cite{benjamin}. Typically the condition for the onset of the deterministic instability can be straightforwardly worked out by means of a traditional linear stability analysis, which constraints the reaction parameters involved in the formulation of the problem \cite{kuramoto,stuart,dipattiBF,mieleBF}. 

Starting from these premises we are here interested in studying the stochastic analog of the BF instability in an open feed-forward topology. In the framework that we shall set to explore, the complex state variable of the CGLE is disturbed by a small exogenous perturbation, 
which configures as an additive white noise, possibly orders of magnitude smaller than the unperturbed oscillation amplitude. 
More specifically, we are interested in assessing the role played by the injected stochastic drive, when the system is operated in a parameters region for which the synchronous limit cycle proves stable under deterministic evolution. As we shall argue, synchronous solutions,  deemed deterministically stable, can turn unstable by agitating the system with an arbitrarily small perturbation. In the case of the CGLE, here assumed as a reference model, we will unfold, and thoroughly characterize, a generalized class of convective instabilities reminiscent of the  BF one. To achieve the sought effect we shall accommodate for  non-normal \cite{trefethen}, diffusive couplings between individual oscillators. In a recent series of papers \cite{nicoletti,Noise driven}, we showed that giant stochastic oscillations, with tunable frequencies, can be obtained, by replicating a minimal model for quasi-cycle along a directed chain of coupled oscillators. Here, the directed link between adjacent oscillators will fuel a self-consistent amplification of the stochastic disturbance, always yielding -- for a sufficiently long chain --  a loss of synchronicity.  Taken all together, our findings constitute a practical example of convective instability~\cite{deissler}, and point to the subtle interplay between noise and topology, capable of changing {\it qualitatively} the system dynamics. This brings into evidence possible failure of a purely deterministic approaches to real life problems.  

The paper is organised as follows: in the next section we will introduce the model to be probed. In particular, we will discuss its synchronized and splay states. We shall then turn to analyze the effect of stochasticity, with reference to the amplification mechanism, as alluded to above. Stochastic non normal patterns are consequently reported to occur, notwithstanding the stability of the homogeneous solution under traditional linear analysis. Finally, we will sum up and draw our conclusions.  

\section{Deterministic Ginzburg-Landau oscillators: synchronized and splay states}
\begin{center}
\begin{figure}[ht]
\centering
\includegraphics[scale=0.6]{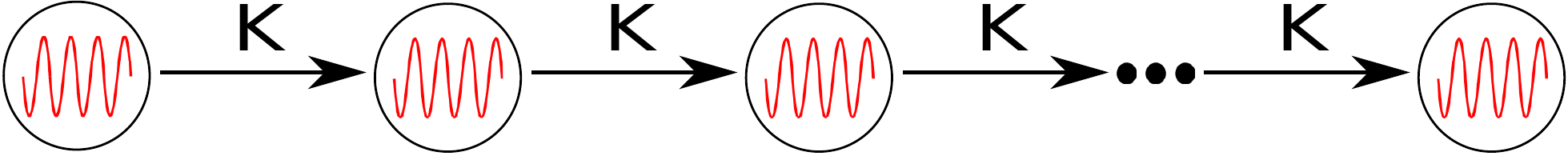} 
\caption{Schematic representation of our system. Each node carries an oscillator and is unidirectionally coupled to its successive neighbour. Parameter $K$ modulates the coupling.}
\label{figuno}
\end{figure}
\end{center}

Our model consists of $\Omega$ diffusively and unidirectionally coupled Ginzburg-Landau oscillators. Each oscillator is described by the complex variable $W_j$ $(1\leq j\leq \Omega)$. The oscillators in this directionally coupled chain (see Fig.\ref{figuno}) 
obey the following ordinary differential equations
\begin{subequations}
\begin{equation}
\frac{\mathrm{d}W_1}{\mathrm{d}t} = W_1 - ( 1 + ic_2)|W_1|^2W_1
\end{equation}
\text{and, for} $j>1$
\begin{equation}
\frac{\mathrm{d}W_j}{\mathrm{d}t}\!\! = \!\! W_j \!-\! ( 1\!+ ic_2)|W_j|^2W_j +(1+ic_1)K (W_{\!j-1}\! -\! W_j)
\end{equation}
\label{dynamic_equation}
\end{subequations}
where $c_1$, $c_2$ are real parameters and $K$ denotes the coupling strength. It is obvious that changing the sign of $K$ and at the same time inverting the boundary conditions is equivalent to reversing the information flow along the chain: therefore in the rest of this paper $K$ is assumed to be positive. The system is also symmetric under the following transformation: $W_j \rightarrow W_j^*$, $(c_1,c_2) \rightarrow - (c_1,c_2)$ which allows us to restrict our focus on half of the $(c_1,c_2)$ parameter plane. Two types of solution are of interest, the \textit{synchronized} and the \textit{splay} ones. The synchronized state (usually denoted as {\sl homogeneous state}, in the vast literature of spatially coupled oscillators) corresponds to the solution 
\begin{equation}
W_j = \exp(- i c_2 t) \quad , \quad  j = 1,\cdots, \Omega.
\label{syncsol}
\end{equation}
By direct inspection of Eq.~(\ref{dynamic_equation}) one can check that any dependence on the spatial coupling $K$ and on the the parameter $c_1$ disappears, and that this solution exists for any value of
$c_2$.

The splay states constitute a family of uniformly rotating solutions with finite constant-in-time phase differences between consecutive nodes. Inserting the general polar form $W_j = \rho_j \exp(-i\theta_j)$ into equation~(\ref{dynamic_equation}), and assuming that $\dot{\rho_j}=0$ and $\theta_j=-c_2 t + \sum^j_{k=1}\phi_k$ (where the $\phi_k = \theta_{k} - \theta_{k-1}$ are constant-in-time phase differences) leads to the recurrence relations
\begin{subequations}
\begin{equation}
\rho_j = \sqrt{\left(1 + K \left[ \frac{\rho_{j-1}}{\rho_j} f(\phi_j) -1 \right]\right)}
\end{equation}
\begin{equation}
0 = c_2 (1-\rho_j^2) + K \left[ \frac{\rho_{j-1}}{\rho_j} g(\phi_j) -c_1 \right]
\end{equation}
\label{recurence relation}
\end{subequations}
where 
\begin{subequations}
\begin{equation}
f(\phi_j) = \cos\phi_j + c_1 \sin\phi_j
\end{equation}
\begin{equation}
g(\phi_j) = c_1\cos\phi_j - \sin\phi_j.
\end{equation}
\label{functions_f_g}
\end{subequations}

\begin{center}
\begin{figure}
\centering
\resizebox{0.7\textwidth}{!}{\includegraphics[scale=1]{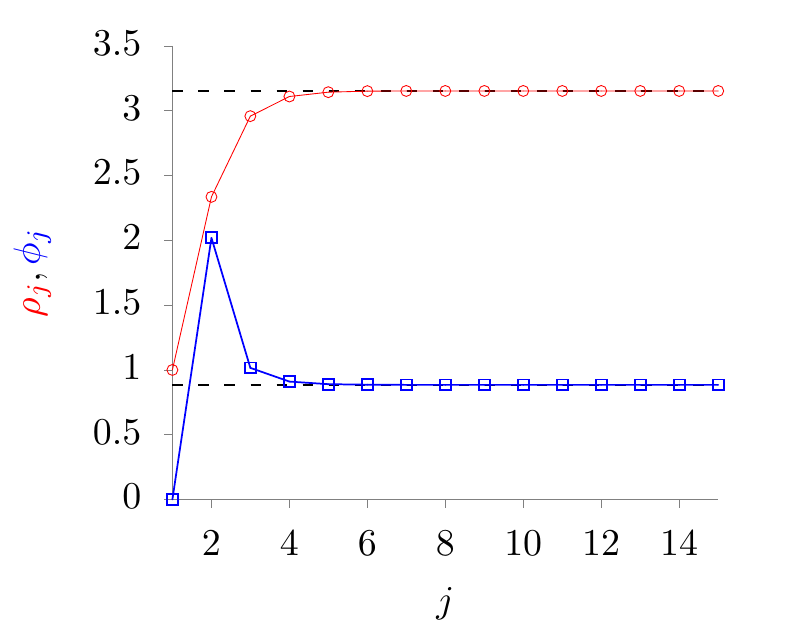}}
\caption{\textsc{Splay state representation} The radius of the limit cycles over the chain $\rho_j$ is depicted by the red solid line, while the blue line stands for the phase difference $(\Mod 2 \pi)$ between two successive nodes. As expected, they converge to the asymptotic values $\rho_{\infty}$, $\phi_{\infty}$ (dashed black lines). The parameters here are $c_1 = -5$, $c_2 = 4$ and $K=4$.}
\label{rho_delta}
\end{figure}
\end{center}

The recurrence relations~(\ref{recurence relation}) need to be completed by a suitable initial condition, where the state variables in the first node of the chain are fixed to some values: we adopt the condition $\rho_1=1$ and $\phi_1=0$, that would be compatible with the synchronous solution (\ref{syncsol}). We avoid reporting explicit calculations, but it can be shown that the recurrence relations beyond the homogeneous solution, also admits a second solution with non zero $\phi_j$, which spatially converges to the asymptotic splay state
\begin{subequations}
\begin{equation}
\label{5a}
\rho_\infty = \sqrt{1 + K \left( f(\phi_{\infty})  -1 \right)}\\
\end{equation}
\begin{equation}
\phi_\infty = 2 \Atan\left[ \frac{1+c_1 c_2}{c_2 - c_1} \right]
\end{equation}
\label{splay_existence}
\end{subequations}
where the special case $\phi_{\infty}=\pm \pi$ occurs in the limit $c_2 \rightarrow c_1$.
In practice, one finds that the asymptotic splay state is rapidly approached along the chain (see Fig.(\ref{rho_delta})) 

The rate of convergence depends on the paramters $K$, $c_1$ and $c_2$,
but here, for the sake of space, we do not report any detailed investigation on this point.

\begin{figure}
\centering
\resizebox{0.7\textwidth}{!}{\includegraphics[scale=1]{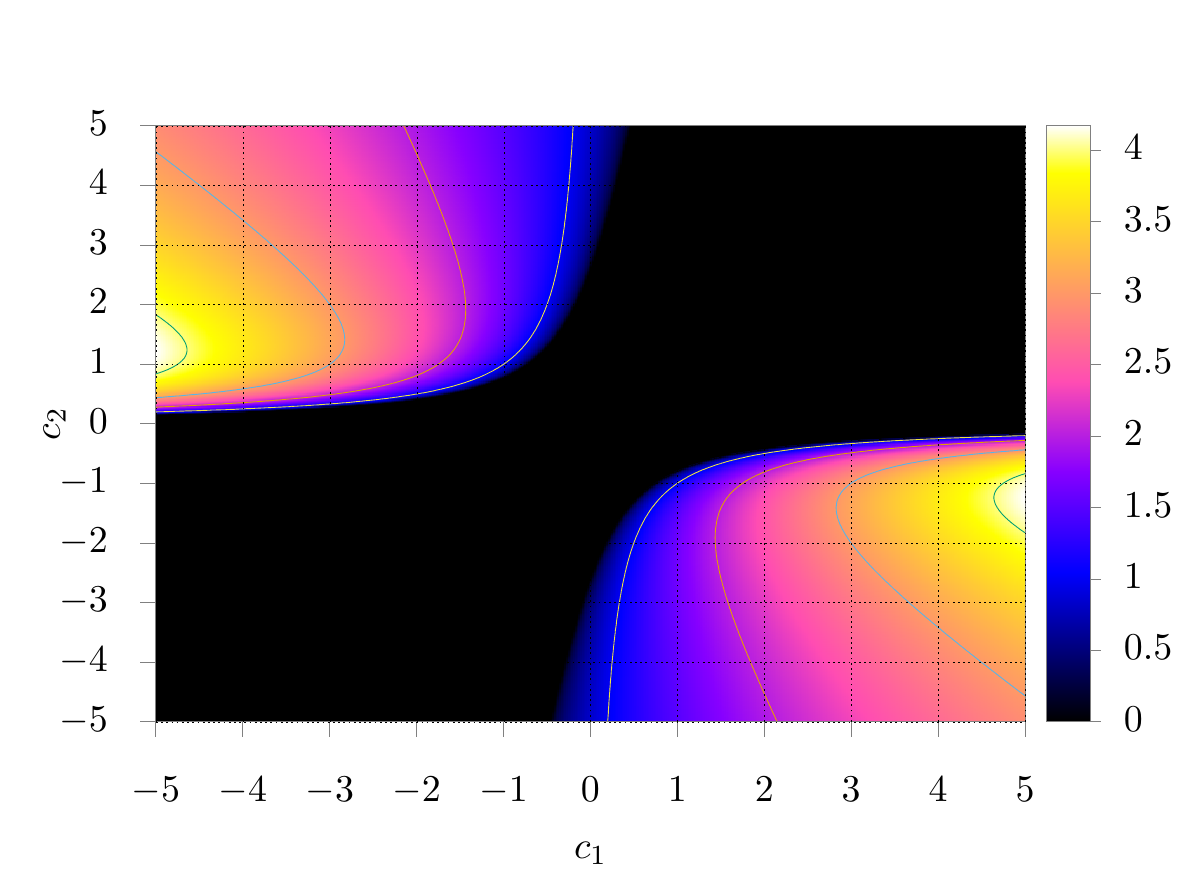}}
\caption{\textsc{Splay state existence:} the color represents the value of $\rho_{\infty}$ (see Eq.(\ref{5a})), while the black zone refers to the region where the splay state does not exist. Here $K=4$.}
\label{att_existence}
\end{figure}

It is important to point out that the existence condition for the splay state is that
$\rho_{\infty}$ is real, i.e. that the argument of the square root in Eq.~(\ref{5a}) is non-negative. As an example, in Fig.~\ref{att_existence} we show 
the region in the $(c_1,c_2)$-plane where the splay state exists for $K=4$: the colour code corresponds to
different positive values of $\rho_\infty$, while the black region indicates where the splay state does not exist. 
As a final remark, we want to point out that there exist an entire family of solutions asymptotically approaching along the chain the splay state (see Eqs.~(\ref{recurence relation})). In these solutions the synchronous state extends to an arbitrary large initial portion of the chain, namely $\rho_j =1$ and $\phi_j =0$ for $j= 1,\cdots, {\bar j}$. For $j > \bar{j}$ phase differences become finite and the solution converges to the asymptotic splay state (5) for large $j$. As we shall discuss later, the existence of this entire family of splay states impacts on the way noise destabilizes the homogeneous synchronized state determining a typical spatio-temporal pattern organization for the stochastic system.

\subsection{Stability of synchronized and splay states}
In order to investigate the stability of the synchronous and of the splay states we 
can perform a standard linear stability analysis. We first introduce small perturbations $\delta\rho_j,\delta\theta_j$ of 
the limit cycles, $W_j = (\rho_j + \delta\rho_j) \exp(i(\theta_j + \delta\theta_j))$ for $1 \leq j \leq \Omega$. Linearizing and retaining the first order in the pertubations leads to an equation that can be put in the general matrix 
form $\delta \dot{\bf v} = {\bf J}({\bf \rho},{\bf \theta}) \delta {\bf v}$ (we adopt the shorthand notation
${\bf (\rho, \phi)} = (\rho_1, \phi_1,\rho_2, \phi_2, \cdots, \rho_\Omega, \phi_\Omega)$), where $\delta {\bf v} = 
\delta(\rho, \phi) $ is the vector 
of perturbations and ${\bf J}$ is the Jacobian matrix associated to dynamics (\ref{dynamic_equation}).
Due to the unidirectional nature of the coupling $K$, ${\bf J}$ exhibits a lower tridiagonal block structure. 
Hence, to assess the stability of any state it is enough to compute the eigenvalues 
$\lambda_{\rho_j}$ and $\lambda_{\theta_j}$ for $1 \leq j \leq \Omega$ of the diagonal $2\times2$ blocks

\begin{subequations}
\begin{equation}
{\bf A}_1 = \left( \begin{array}{lr}
-2 & \;\;0\\
&\\
-2 c_2 &\;\; 0
\end{array}\right)
\label{block_1}
\end{equation}
\text{and} 
\begin{equation}
{\bf A}_j \!\!= \!\!\left( \begin{array}{ll}
\!(1 - 3 \rho_j^2 - K) & \;\;K \rho_{j-1} g (\phi_j)\\
&\\
\!-\!\left( 2 c_2 \rho_j +K\frac{\rho_{j-1}}{\rho_j} g (\phi_j) \right)
                     &\;\; -K \frac{\rho_{j-1}}{\rho_j} f (\phi_j)
\end{array}\!\!\right)
\label{block_j}
\end{equation}
\end{subequations} 
for $2 \leq j\leq \Omega$.

The eigenvalues of the first block $A_1$ are $\lambda_{\rho_1} \!\!=\!\! -2$ and $\lambda_{\theta_1}\!\!=\!\!0$, 
the latter reflecting marginal stability towards global phase rotations. A given limit cycle solution is stable only if the 
all other block complex eigenvalues have a negative real part, i.e. $\Re(\lambda_{\rho_j}) <0$ and $\Re(\lambda_{\theta_j}) <0$ 
for $ 2\leq j \leq \Omega$.

\begin{center}
\begin{figure}
\centering
\resizebox{0.7\textwidth}{!}{\includegraphics[scale=1]{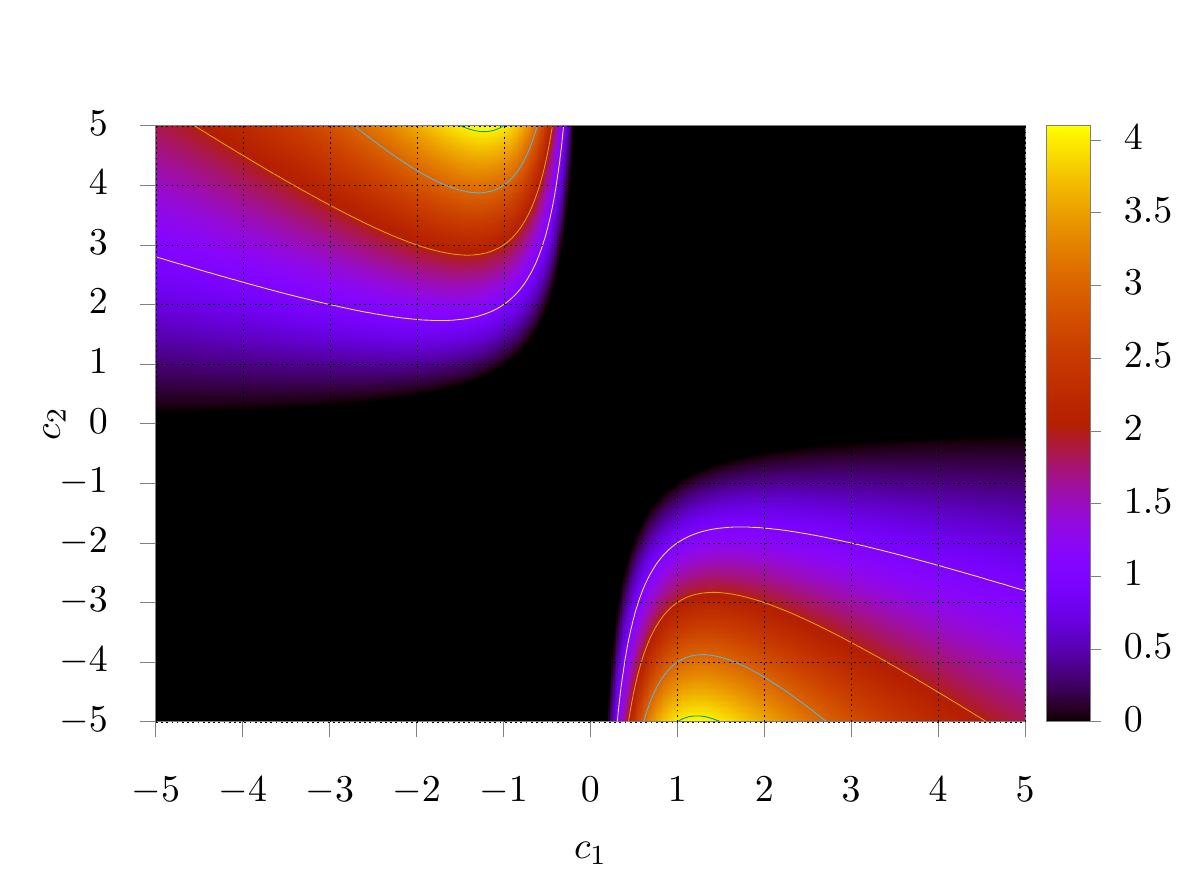}}
\caption{\textsc{Synchronized state stability:} On this panel we display the value of $K^H_{min}$ (see Eq.~(\ref{stab_homogeneous})) beyond which the synchronized state is stable.  In the black region $1+c_1 c_2 \geq 0$ and the synchronized state is always stable.}
\label{K_min}
\end{figure}
\end{center}

The synchronized state, where $\rho_j=1$, $\phi_j = 0 \quad \forall j$, is stable independently of $K$ for $1+c_1 c_2 \ge 0$, while for $1 + c_1 c_2 <0$ only if the following condition holds:
\begin{equation}
K > K^H_{min} = -\frac{2(1+ c_1c_2)}{1 + c_1^2}.
\label{stab_homogeneous}
\end{equation} 
Therefore, for each couple $(c_1,c_2)$ we can find a minimum coupling value $K^H_{min}$ such that the synchronized state is stable. The resulting stability map is shown in Fig.~\ref{K_min}. Notice that the condition $1+c_1c_2 < 0$ is sufficient for the onset of the instability, when the CGLE is defined on a continuous spatial support \cite{dipattiBF}. In fact, this is known as the condition of the Benjamin-Feir instability for the CGLE \cite{benjamin,dipattiBF}.

\begin{center}
\begin{figure}
\centering
\resizebox{0.7\textwidth}{!}{\includegraphics[scale=1]{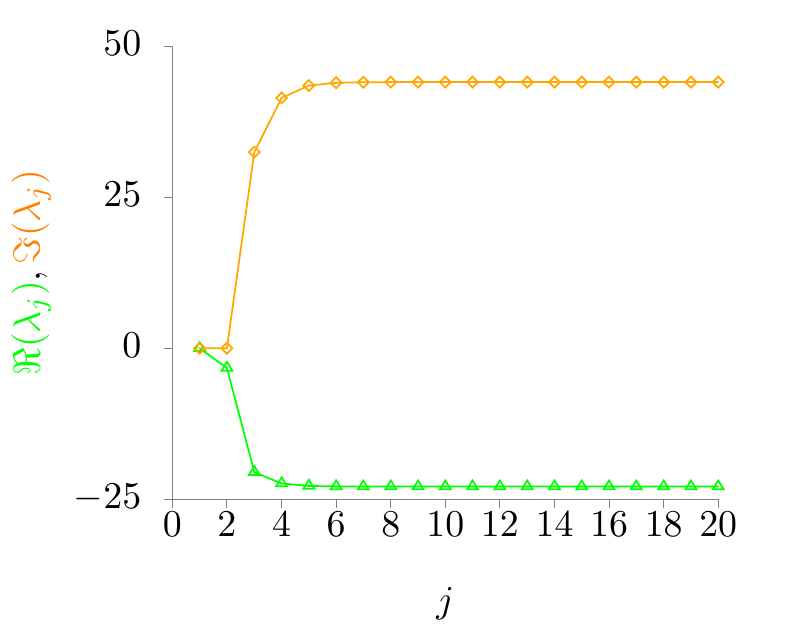}}
\caption{\textsc{Splay state linear stability analysis} The green curve represents the real part of the largest eigenvalue for each node $j$ while the yellow line is the corresponding imaginary part. The parameters here are $c_1 = -5$, $c_2 = 4$ and $K=4$. In this example, the splay state is characterized by $\bar{j}=1$.}
\label{eig_stab}
\end{figure}
\end{center}

\begin{center}
\begin{figure}
\centering
\resizebox{0.75\textwidth}{!}{\includegraphics[scale=1]{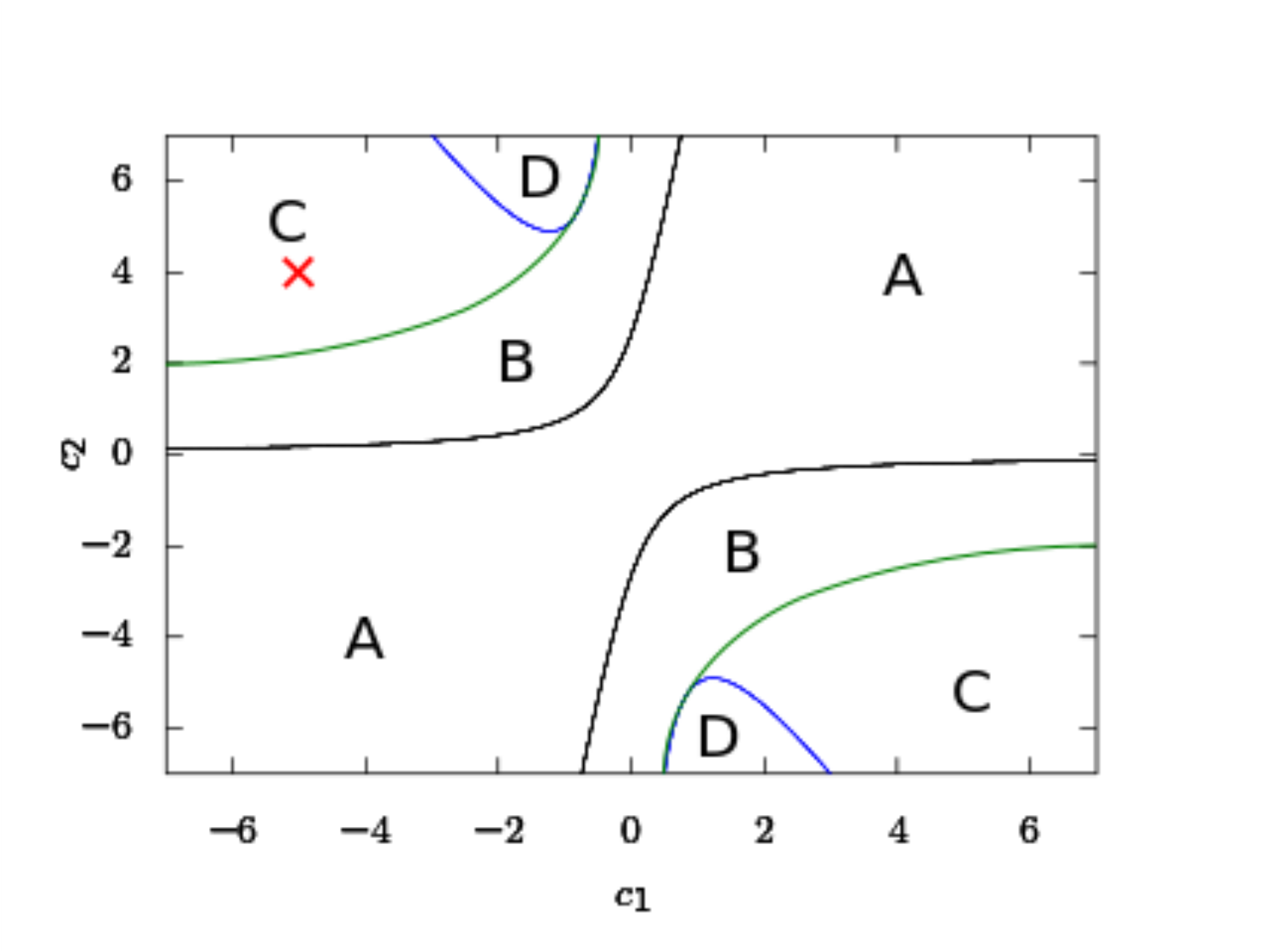}}
\caption{Diagram of the existence and stability of the synchronized and of the splay states for $K=4$: in region A, whose boudaries are fixed by the condition $\rho_{\infty} =0$, only the synchronized state exists and is stable; in region B both 
states exist, but only the synchronized one is stable; in region C both states are stable, while in region D the splay state only is stable. 
}
\label{map_stability}
\end{figure}
\end{center}

Stability analysis is more complicated for the splay state. Making use of the recurrence relations (\ref{recurence relation})
we can first compute $\rho_j$ and $\theta_j$ to evaluate the Jacobian blocks ${\bf A}_j$ (see Eqs.~(\ref{block_j})). Then we can
assess the stability of the splay state in the plane $(c_1,c_2)$ by computing the Jacobian matrix eigenvalues.
An example of the outcome of this procedure is shown in Fig.~\ref{eig_stab}, where the parameters have been set to the values $c_1= - 5$, $c_2=4$ and $K=4$. Here all eigenvalues for $j>1$ have a negative real part, so that the splay state is linearly stable. Notice the fast convergence of the eigenvalues to their asymptotic state values. 

The analysis of the synchronized state and of the splay one of the directed chain of coupled CGL oscilators is summarized in
Fig.~\ref{map_stability} for the case $K=4$. The different regions of this diagram are described in the caption; the red cross locates the point in the diagram which defines our working condition as selected in the forthcoming sections when investigating the stochastic version of the directed chain of coupled CGL oscillators. More details on linear stability analysis are given in~\ref{appendix_stability}.

\section{Effects of stochasticity}
\subsection{Linear amplification mechanism}
\begin{center}
\begin{figure}
\centering
\resizebox{0.7\textwidth}{!}{\includegraphics[scale=1]{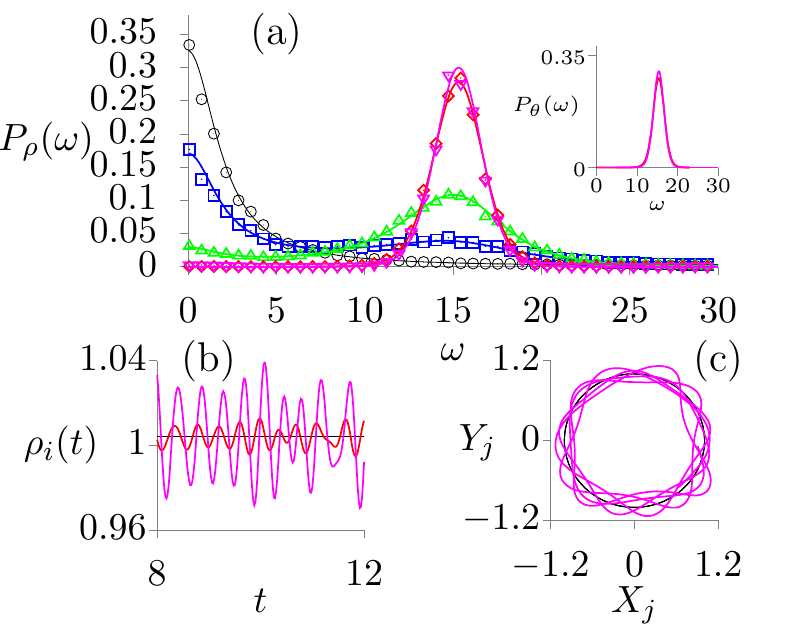}}
\caption{(a) Normalized power spectra of different nodes along the chain: the solid lines stands for the theoretical calculation while the symbols correspond to  numerically computed power spectra using the Euler-Maruyama algorithm. The displayed agreement confirms the validity of the analytic calculations. In the inset: normalized power spectra for longitudinal fluctuations. (b) Trajectories of $\rho_j$. The amplification phenomenon can be clearly appreciated. (c) Phase portrait of $\left(X_j(t),Y_j(t)\right)$ where $X_j$ and $Y_j$ respectively stand for the real and the imaginary part of the complex variable $W_j$. Oscillations extend along the radial direction, and progressively alter the unperturbed limit cycle profile. 
 The parameters here are $c_1 = -5$, $c_2 = 4$, $\sigma=10^{-5}$ and $K=4$. Each color designs a specfic node: 1 black, 2 blue, 3 green, 8 red, 9 violet.}
\label{power_spectrum_fig}
\end{figure}
\end{center}

The stochastic version of the deterministic model (\ref{dynamic_equation}) reads
\begin{equation}
\frac{\mathrm{d}W_j}{\mathrm{d}t}\!\!=\!\! W_j - ( 1 + ic_2)|W_j|^2W_j +(1+ic_1)K (W_{j-1}\!-\!W_j) + \sigma \eta_j(t)
\label{stochastic_dynamic_equation}
\end{equation}
where $\sigma$ is the noise amplitude, $\eta_j = \Re(\eta_j) + i \Im(\eta_j)$ is a complex
additive noise with zero mean and correlators $\braket{\Re(\eta_j)(t)\Re(\eta_l)(t')} =
\braket{\Im(\eta_j)(t)\Im(\eta_l)(t')}=
\delta_{jl}\delta(t\!-\!t')$.
In what follows the numerical investigations of the stochastic dynamics (\ref{stochastic_dynamic_equation})
has been performed for the parameter values $(c_1,c_2,K) = (-5,4,4)$ (see the red cross in
Fig. (\ref{map_stability}) ), where both the synchronized and the splay state of the deterministic dynamics
are linearly stable. We want to investigate the effects 
of a small additive noise on the deterministic evolution (\ref{dynamic_equation})~\cite{asslani2,mckane1,mckane2}. 
In practice, we have always taken $\sigma=10^{-5}$, a value which is five orders of magnitude smaller than the 
oscillations amplitude of the synchronized state. As shown in~\ref{nature_noise}, 
Equation~(\ref{stochastic_dynamic_equation}) can be rewritten for the polar components of the complex
variable $W_j$, while the corresponding noise components remain delta-correlated and -- at least near the
limit cycle solutions -- additive.
In practice, we have studied the effects of the noise--induced fluctuations around these states. We know from the previous 
section that both deterministic states are indeed stable limit cycles with a complex eigenvalues Jacobian. This guarantees the presence of stochastic oscillations, also called quasi-cycles~\cite{mckane3, mckane4}, on the top of the deterministic
stable states. Then, we can proceed to the Fourier analysis of our system linearized around each limit cycle. We denote by ${\bf \delta\tilde{v}}$ and ${\bf \tilde{\xi}}$ the Fourier transforms of the perturbations vector ${\bf \delta v}$ and of the polar white noise $\xi \equiv (\xi_\rho, \xi_\theta)$, respectively. We can readily obtain ${\bf \delta\tilde{v}}_j \!\!=\!\! \sum^{2 \Omega}_{l=1} \Phi_{jl}^{-1}(\omega){\bf \tilde{\xi}}_l$, where $\Phi_{jl}\!=\!-J_{jl}- i \omega \delta_{jl}$. To pursue the analysis of the oscillations we compute the power spectrum density matrix of the fluctuations in the vicinty of the attractor~\cite{challenger} \vspace{-0.2cm}
\begin{equation}
\braket{{\bf \delta\tilde{v}}_l(\omega) {\bf \delta\tilde{v}}_j(\omega)}\!\!=\!\!P_{lj}(\omega)\!\!=\!\! \sum^{2 \Omega}_{k=1} \Phi_{lk}^{-1}(\omega)\!\!\left({\Phi^\dagger}_{kj}\right)^{\!-1}\!\!(\omega).
\label{power_spectrum}
\end{equation}
Its diagonal entries are the power spectrum of transversal ($j$ odd) and longitudinal ($j$ even) oscillations around both solutions. We first focus on the transversal, radial, fluctuations around the synchronized state. In Fig.~\ref{power_spectrum_fig}(a) we depict the power spectrum of several nodes. The solid line stands for the analytical power spectrum computed from equation~(\ref{power_spectrum}) while symbols correspond to direct numerical simulations of equation~(\ref{stochastic_dynamic_equation}), using the Euler-Maruyama algorithm ($\mathrm{d}t = 0.001$). The power spectrum of the first node, peaked at zero frequency (circle, black line) is  the one of white noise. As we proceed along the chain, the peak of the power spectrum progrssively
shifts towards higher frequencies. The profiles around the peak become narrower (thus singling out a well defined
oscillation frequency), while fluctuations are amplified along the chain. This amplification can be well appreciated by
direct  inspection of Fig.~\ref{power_spectrum_fig}(b). Such amplification and modulation proceeds along the chain as long as the linear approximations holds. Out of this approximation, nonlinear effects should take over and stop the amplification process. Note that an analogous phenomenon was already discussed in~\cite{Noise driven} for noisy fluctuations around a single fixed point. Since the structure of the Jacobian remains essentially the same for the splay state, here we face a qualitatively identical situation. A similar amplification mechanism takes place for longitudinal fluctuations around
 both stable states, as exemplified in the inset of Fig.~\ref{power_spectrum_fig}(a). However, longitudinal oscillations are typically characterized by a broader spectrum, possibly due to the softer nature of the phase direction with respect to the radial one for Ginzburg-Landau potentials. To summarize our findings, noisy fluctuations around both attractors are amplified and modulated as one proceeds along the chain to yield sharper and stronger oscillations. While nonlinear effects would eventually arrest this amplification process, the linear mechanism is typically enough to overcome the attractor linear stability itself. These features are mainly due to the unidirectional structure of the Jacobian, which is highly non-normal. It is well known that non-normality amplifies transient dynamics~\cite{Caswell,Lloyd, Allee} and may lead to convective instability~\cite{deissler}. Here the presence of noise makes this amplification perpetual~\cite{nicoletti}. 

\subsection{Pattern formation}
\begin{center}
\begin{figure}[h!]
\centering
\resizebox{0.7\textwidth}{!}{\includegraphics[scale=1]{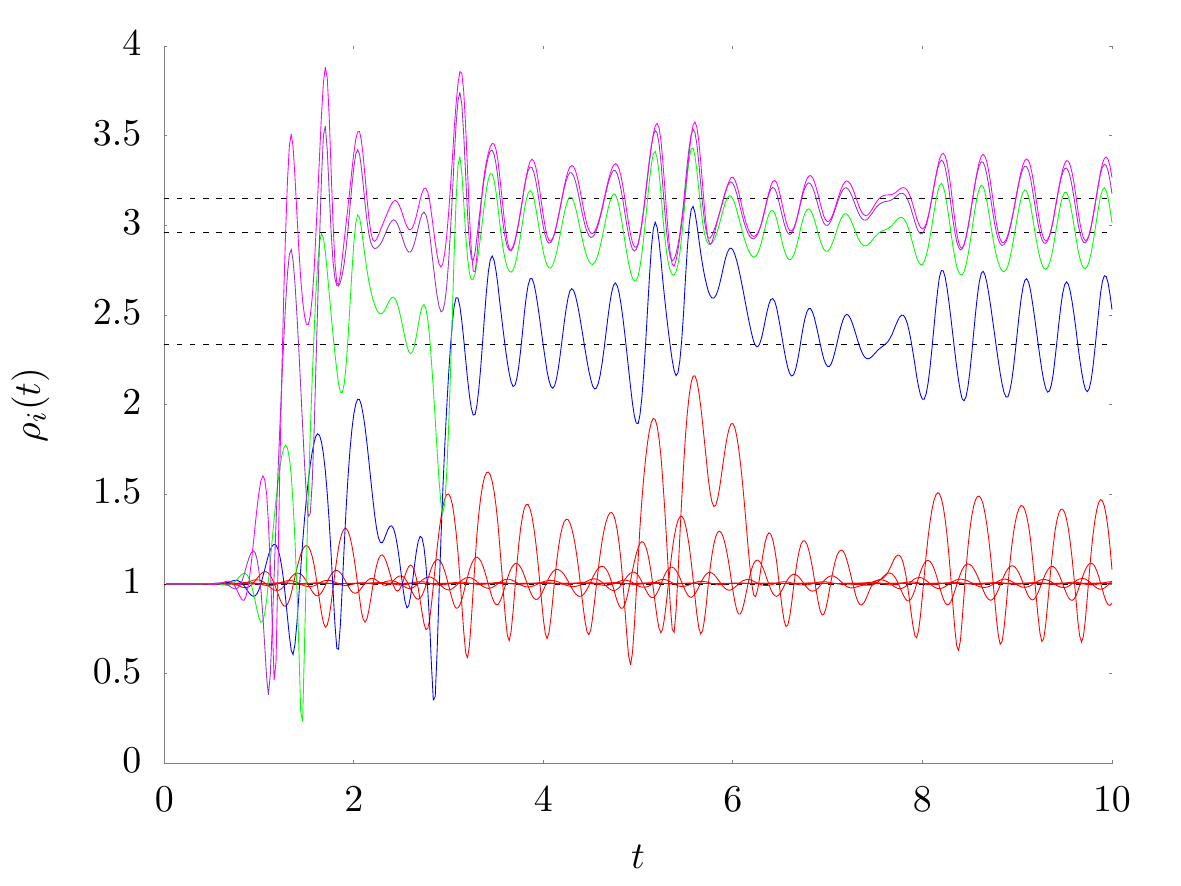}}
\caption{Time evolution for selected amplitude $\rho_j(t)$. Each solid line refers to the time series of $\rho_j(t)$. Each color defines a specific group, in red are depicted all the nodes that remain on the synchronized attractor (from the first to the $9^{th}$ node). The $10^{th}$ node (blue line) is the first able to escape from the homogenous attractor. The successive nodes ($11^{th}$ green and $12^{th}$ violet) converge progessively to the asymptotic value $\rho_{\infty}$ of the splay state. The parameters here are $c_1 = -5$, $c_2 = 4$, $\sigma=10^{-5}$ and $K=4$.}
\label{transient}
\end{figure}
\end{center}

\begin{center}
\begin{figure}
\centering
\resizebox{0.7\textwidth}{!}{\includegraphics[scale=1]{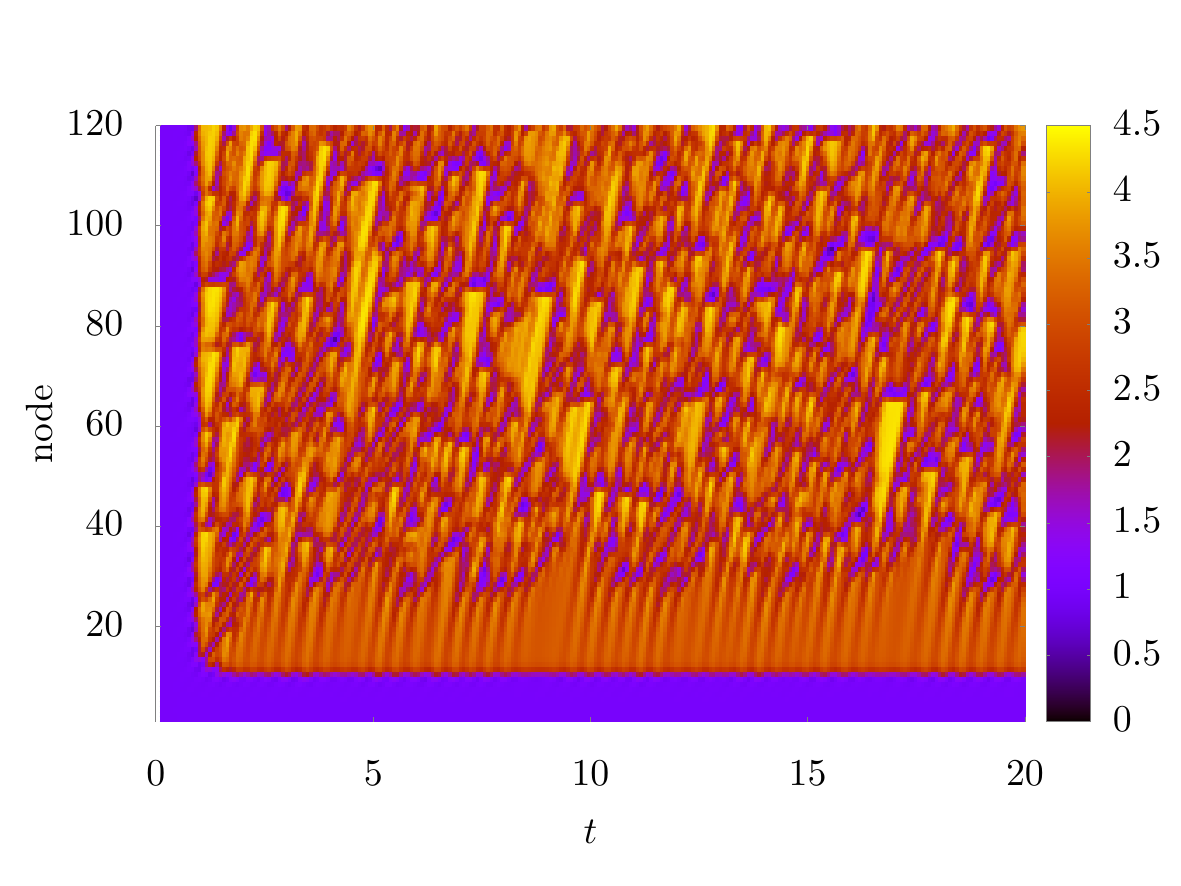}}
\caption{Typical spatiotemporal pattern of our system, the 'space' (nodes) is the $y-$axis while time is in abscissa. One
can easily recognize the transient in which all the nodes are int the synchronized state. The orange plateau stands for the splay sta e (node $10 \rightarrow 30$) and precedes the blurred region, where system erratically jumps from one state
to another. The parameters here are $c_1 = -5$, $c_2 = 4$, $\sigma=10^{-5}$ and $K=4$.}
\label{spatiotemporal}
\end{figure}
\end{center}

Why is this so important? Let's immagine the following scenario where both solutions exist and are stable. We then seed the following initial conditions $\rho_j(t=0) = 1$, $\phi_j(t=0) = 0$ all over the chain. What we expect from a naive linear stability analysis is that, for small noise amplitudes, the system will remain in the vicinity of the synchronized state, with fluctuations of the order the noise amplitude $\sigma$. On the contrary, our analysis reveals that the amplification mechanism here discussed will drive  the system to progressively explore larger portions of the available phase space, until it eventually reaches the splay state. This is illustrated in Fig.~\ref{transient}, where we show the radial time series of successive nodes. The time series of the first nodes are plotted in red: they remain settled on the synchronized state, the amplification on these first nodes not being strong enough to escape from its basin of attraction. After the $10^{th}$ node (blue line) fluctuations are now strong enough to escape, and reach the second attractor, settling on the splay state radius $\rho_{\bar{j}+1}$. Nodes to the right converge to successive radii $\rho_{j}$ with $j>\bar{j}$. The attractor values $\rho_j$, each represented by a dashed line, is found thanks to the recurence relations~(\ref{recurence relation}). They are in good agreement with the time series simulations performed by an Euler-Maruyama algorithm. Obviously, this could not be expected from a traditional linear stability analysis.
Actually, by careful observation of the time series, one can realize that these shifts from the synchronized to the splay state take place as a sort of zipping mechanism from right to left. The rightmost nodes display larger oscillations, and are the first ones to escape the synchronized state (violet lines in Fig.~\ref{transient}). Progressively, the escape point from the synchronized state moves to the left. In this process, individual time series typically jump synchronously from one value $\rho_{\bar{j}+k}$ to the next $\rho_{\bar{j}+k+1}$ (see the green and blue timeseries in Fig.~\ref{transient}).
This zipping process continues towards the left, until the oscillations on node $\bar{j}$ (the $9^{th}$ node in our case) become strong enough to escape the synchronized part of the attractor, where $\rho=1$. \newline
A direct consequence of this mechanism is the formation of spatiotemporal patterns~\cite{mckane1,asslani, Tomaso} as shown in Fig.~\ref{spatiotemporal}. Our system is initially prepared on the synchronized state and exposed to a noise of amplitude $\sigma=10^{-5}$. After some time we see that the rightmost nodes easily reach the second attractor. However, as we already discussed, the same amplification and modulation mechanism holds on the splay state. The fluctuations therefore keep on being amplified along the chain allowing the rightmost nodes of our system to travel erratically in phase space. This is exemplified by the blurred part of Fig.~\ref{spatiotemporal}. Here the mechanism of desynchronization is quite obvious, being the combination of two ingredients: noise and non-normality. While noise is needed to inject some dynamics in the otherwise stable limit cycle, the non-normality is essential to amplify these fluctuations. This is what makes the system deviate from the synchronized to the splay state and then enter an erratic dynamics.
\section{Conclusion}

Noise is often unavoidable and, as such, it should be accommodated for in realistic models of complex natural phenomena. A particularly interesting setting is faced when the stochastic perturbation, being it of endogenous or exogenous origin, resonates with the degree of inherent non-normality. This situation, as displayed by the examined system, yields a self-consistent amplification of the noise component at short times. The resulting growth of the perturbation can in fact drive a symmetry breaking instability, for a choice of the parameters that would instead result in a stable deterministic evolution. In order to dig into this question, we have here examined a directed chain of diffusively coupled, Ginzburg-Landau oscillators. Oscillators are shaked by an external fluctuating drive, of arbitrarily small strength. The system is initiated in a region of parameters where the synchronous solution proves stable, under the deterministic scenario. Working in this setting, we provided analytical and numerical evidence for a noise induced instability which follows the self-consistent amplification of the imposed disturbance across the chain. The limit cycles get modulated along the transversal direction: almost regular, radial oscillations are displayed, which gain in potency node after node. When the transversal modulation gets large enough, oscillators escape the basin of attraction of the synchronized solution, visiting  a non trivial attractor, that we have analytically characterized. The interaction between the two attractors yield complex emerging patterns reminiscent of the deterministic Benjamin-Feir instability.    
The combination of noise and asymmetric couplings can radically alter the limit cycle dynamics: bistability and associated patterns rise, as the noisy signal is dynamically processed, along the unidirectionally coupled chain. It is worth mentioning that an analogous behavior, due to the forward amplification mechanism, is also expected when the (arbitrarely small) noise is only injected in the leftmost node and not on all degrees of freedom as in our current setup. Traditional (deterministic) linear stability analysis is unable to grasp the essence of the phenomenon, an observation which we find particularly relevant given the recent reports on the ubiquity of non-normality in real systems, from communication networks to foodwebs~\cite{timoteo}. More refined approaches, such as convective Lyapunov exponents~\cite{kaneko, lepri1, lepri2, jiotsa} should however be able to predict a convective instability at the purely deterministic level. 
Resilience to synchronization might prove a valuable asset, exploited to oppose the onset of pathological states, as e.g.  epileptic seizures in brain dynamics. Future investigations are planned to shed light onto these families of noise--instigated instabilities, assisted by the non-normal topology of the underlying support, beyond the simplistic case study here considered. 

\section*{Acknowledgments}
The authors acknowledge financial support from H2020-MSCA-ITN-2015 project COSMOS  642563. We thank Arkady Pikovsky for useful comments.

\appendix

\section{Linear stability analysis}
\label{appendix_stability}
We now give more details on the linear stability analysis. Consider small perturbations $\delta \rho_j (t) \ll 1$ and
$\delta \theta_j (t) \ll 1$ of the limit cycle solutions, $W_j =
(\rho_j + \delta \rho_j) e^{i (\theta_j + \delta
  \theta_j)}$. Linearizing we obtain to first order in the perturbations
\begin{eqnarray}
\label{eq:linear0A}
\delta \dot{\rho}_1 &=& - 2 \delta \rho_1 \\
\label{eq:linear0B}
\delta \dot{\theta}_1 &=& - 2 c_2 \delta \rho_1 
\end{eqnarray}
and for $j > 1$

\begin{eqnarray}
\label{eq:linearjA}
\delta \dot{\rho}_j &=& \delta \rho_j \left[1 - 3 \rho_j^2 - K \right] + \delta \rho_{j-1} K f(\phi_j) +   \left( \delta \theta_j -\delta \theta_{j-1}\right) K \rho_{j-1}g(\phi_j) \\
\label{eq:linearjB}
\delta \dot{\theta}_j &=& \delta \rho_j \left[ - 2 c_2  \rho_j +K \frac{\rho_{j-1}}{\rho_j} g(\phi_j)\right]
                                   + \delta \rho_{j-1} \frac{K}{\rho_j}g(\phi_j) \nonumber\\
                                   &&- \left( \delta \theta_j - \delta\theta_{j-1}\right) K\frac{\rho_{j-1}}{\rho_j}f(\phi_j) 
\end{eqnarray}
where the $\rho_j$ and $\phi_j$ need to be evaluated on either the
synchronized or the splay state attractor. Obviously zeroth order
terms stemming from the linearization procedure vanish by construction
when evaluated on these two attractors.\\

Rewriting the linearized equations in a matrix form highlights their
simple block structure, due to the unidirectional input from one node
to the next. We introduce the $2\Omega$ dimensional perturbation vector
$\delta {\bf v} \equiv (\delta \rho_1, \delta \theta_1, \delta \rho_2,
\delta \theta_2, \ldots ,\delta \rho_\Omega, \delta \theta_\Omega)^T$ and write
\begin{equation}
\delta \dot{\bf v} = {\bf J}   \, \delta {\bf v}
\end{equation}
where the Jacobian ${\bf J}$ is a $2\Omega \times 2\Omega$ lower tridiagonal block matrix, composed of $2 \times 2$ blocks that describe the in-node linearized dynamics (${\bf A}$ matrices in the following) or the (linearized) interaction with the previous node (${\bf B}$ matrices).\\

For instance, in the case of the synchronized state one has

\begin{equation}
{\bf J}_H = 
\left( \begin{array}{ccccc}
{\bf A}_1 & {\bf 0} & {\bf 0} &  {\bf 0} & \ldots \\
{\bf B}_H & {\bf A}_H & {\bf 0}  & {\bf 0} & \ldots \\
{\bf 0} & {\bf B}_H & {\bf A}_H & {\bf 0} & \ldots \\
{\bf 0} & {\bf 0} & {\bf B}_H & {\bf A}_H & \ldots \\
\vdots & \vdots & \vdots & \vdots & \ddots
\end{array}\right)
\end{equation}

where 
\begin{equation}
{\bf A}_1 = \left( \begin{array}{lr}
-2 & 0\\
&\\
-2 c_2 & 0
\end{array}\right)
\end{equation} 
describes the stability of the first uncoupled Landau-Stuart node,
while
\begin{equation}
{\bf A}_H = \left( \begin{array}{lr}
-(2+K) & \;\;K c_1\\
&\\
-(2 c_2 +K c_1)  &\;\; -K
\end{array}\right)
\;,\;
{\bf B}_H = \left( \begin{array}{lr}
K & -K c_1\\
&\\
K c_1 & K
\end{array}\right)
\end{equation} 
originate from the other nodes $(j>1)$.\\

Using simple block matrices results one can show that 
\begin{equation}
\det\left({\bf J}_H - \lambda {\bf I}_{2\Omega} \right) = \det\left({\bf
A}_1 - \lambda {\bf I}_{2} \right) \, \left[\det\left({\bf A}_H - \lambda {\bf I}_{2} \right)\right]^{\Omega-1}
\end{equation}
(where  ${\bf I}_h$ is the $h \times h$ identity matrix)
so that the eigenvalues of ${\bf J}_H$ are given by the ones of ${\bf
A}_1$ and the ones of ${\bf A}_H$ (with multiplicity $\Omega-1$).\\
We easily verifiy that ${\bf A}_1$ has eigenvalues $\lambda_\rho = -2$ and $\lambda_\theta = 0$ and consequently is stable. We are therefore interested in the eigenvalues $\lambda_H$ of ${\bf A}_H$ that give
\begin{equation}
\lambda_H^\pm = -1 - K \pm \sqrt{1 - 2 K c_1 c_2  - K^2 c_1^2}
\label{a10}
\end{equation}
The real part of the largest eigenvalue $\lambda_H^+$ has two zeros for $K=0$ and $K=K^H_{min}$ with 
\begin{equation}
K^H_{min} = -\frac{2(1+ c_2c_1)}{1 + c_1^2}
\end{equation} 
We can determine the stability condition $\mathcal{R} \left[\lambda_H^+ \right] < 0$ by the sign of $K^H_{min}$ and of the small $K$ expansion of equation (\ref{a10}),
\begin{equation}
\lambda_H^+ \approx - K (1 + c_1 c_2) < 0
\end{equation},
which gives the sign of the $K$ derivative of $\mathcal{R} \left[\lambda_H^+ \right]$ near $K=0$. Note that they are both controlled by the sign of $1 +c_1c_2$, so that one immediatly obtains the homogeneous state stability condition given in the main text.

The splay states give rise to a slightly more complicated Jacobian
matrices ${\bf J}_S^{(\bar{j})}$. The first $\bar{j}$ rows are identical to
the ones of ${\bf J}_H$, while the following ones are obtained
evaluating the linearized equation along the splay state part of the
attractor. For instance, for $\bar{j}=2$ we have
 
\begin{equation}
{\bf J}_S^{(2)} = 
\left( \begin{array}{ccccc}
{\bf A}_1 & {\bf 0} & {\bf 0} &  {\bf 0} & \ldots \\
{\bf B}_H & {\bf A}_H & {\bf 0}  & {\bf 0} & \ldots \\
{\bf 0} & {\bf B}_3 & {\bf A}_3 & {\bf 0} & \ldots \\
{\bf 0} & {\bf 0} & {\bf B}_4 & {\bf A}_4 & \ldots \\
\vdots & \vdots & \vdots & \vdots & \ddots
\end{array}\right)
\end{equation}

with
\begin{equation}
{\bf A}_j = \left( \begin{array}{lr}
(1 - 3 \rho_j^2 - K) & \;\;K \rho_{j-1} g (\phi_j)\\
&\\
-\left( 2 c_2 \rho_j +K\frac{\rho_{j-1}}{\rho_j} g (\phi_j) \right)
                     &\;\; -K \frac{\rho_{j-1}}{\rho_j} f (\phi_j)
\end{array}\right)
\end{equation} 
and 
\begin{equation}
{\bf B}_j = \left( \begin{array}{lr}
K f(\phi_j) &\;\;\; -K \rho_j g(\phi_j)\\
&\\
\frac{K}{\rho_j} g (\phi_j) &\;\;\; K \frac{\rho_{j-1}}{\rho_j}  f(\phi_j) 
\end{array}\right)
\end{equation} 
where functions $g$ and $f$ are defined in the main text in equations~(\ref{functions_f_g}).
Obviusly, for $j \gg \bar{j}$ we have 
\begin{equation}
{\bf A}_j \approx {\bf A}_\infty = \left( \begin{array}{lr}
(1 - 3 \rho_\infty^2 - K) & \;\;K \rho_\infty g (\phi_\infty)\\
&\\
-\left( 2 c_2 \rho_\infty +K g (\phi_\infty) \right)
                     &\;\; -K f (\phi_\infty)
\end{array}\right)
\end{equation} 
and
\begin{equation}
{\bf B}_j \approx {\bf B}_\infty = \left( \begin{array}{lr}
K f(\phi_\infty) &\;\;\; -K \rho_\infty g(\phi_\infty)\\
&\\
\frac{K}{\rho_\infty} g (\phi_\infty) &\;\;\; K  f(\phi_\infty) 
\end{array}\right)
\end{equation} 

Once again we have
\begin{eqnarray}
\det\left({\bf J}_S^{(\bar{j})} - \lambda {\bf I}_{2\Omega} \right) &\!\!=\!\!& \det\left({\bf
A}_1 - \lambda {\bf I}_{2} \right) \, \left[\det\left({\bf A}_H -
\lambda {\bf I}_{2} \right)\right]^{\bar{j}-1}  \nonumber\\  
                                        &&\prod_{j=\bar{j}+1}^N \left[\det\left({\bf A}_j -
\lambda {\bf I}_{2} \right)\right]
\end{eqnarray}
so that to estimate the eigenvalues of ${\bf J}_S^{(\bar{j})}$ we
also need to compute the eigenvalues of the matrices ${\bf A}_j$,
evaluated on the splay attractor values $\rho_j$ and $\phi_j$ obtained
from the recurrence equations~(\ref{recurence relation}).

\section{Nature of the noise}
\label{nature_noise}
In this appendix we shall demonstrate that the additive stochastic corrections we introduced in our system (see equation~(\ref{stochastic_dynamic_equation})) remains of the same kind in polar form. We first write the ordinary differential equations for the real and imaginary part of $W_j=X_j + i Y_j$. After few lines of algebra we end up with
\begin{eqnarray}
\frac{\mathrm{d}X_j}{\mathrm{d}t} &\!\!=\!\!& X_j + \left( X_j^2 + Y_j^2 \right) \left( - X_j + c_2 Y_j \right) \nonumber\\  
                                        && + K \left( X_{j-1} - X_j + c_1 \left( Y_j - Y_{j-1} \right) \right)  \nonumber\\  && + \sigma \eta_j^X\\
\frac{\mathrm{d}Y_j}{\mathrm{d}t} &\!\!=\!\!& X_j + \left( X_j^2 + Y_j^2 \right) \left( - c_2 X_j - Y_j \right) \nonumber\\  
                                        && + K \left( Y_{j-1} - Y_j + c_1 \left( X_{j-1} - X_{j} \right) \right)  \nonumber\\  && + \sigma \eta_j^Y\\ \nonumber
\end{eqnarray}
Writing in polar form $W_j\!\! =\!\! \rho_j \exp(i\theta_j)$ implies that $\rho_j \!\!=\!\! \sqrt{X_j^2 + Y_j^2}$ and $\theta_j\!\! =\!\! \Atan \left( \frac{Y_j}{X_j} \right)$. In terms of O.D.E.s it means that 
\begin{subequations}
\begin{equation}
\rho_j \frac{\mathrm{d}\rho_j}{\mathrm{d}t} = X_j \frac{\mathrm{d}X_j}{\mathrm{d}t} + Y_j \frac{\mathrm{d}Y_j}{\mathrm{d}t}
\end{equation}
\begin{equation}
\rho_j^2 \frac{\mathrm{d}\theta_j}{\mathrm{d}t} = X_j \frac{\mathrm{d}Y_j}{\mathrm{d}t} - Y_j \frac{\mathrm{d}X_j}{\mathrm{d}t}
\end{equation}
\end{subequations}
We now want to obtain the Langevin equation for $\rho_j$ and $\theta_j$, this leads to 
\begin{eqnarray}
\rho_j \frac{\mathrm{d}\rho_j}{\mathrm{d}t} &\!\!=\!\!& \rho_j^2 - \rho_j^4 + K \left( -\rho_j^2 + \rho_j\rho_{j-1} f(\phi_j) \right)  + \sigma \left( X_j \eta_j^X + Y_j \eta_j^Y \right)\\
                                        \rho_j^2 \frac{\mathrm{d}\theta_j}{\mathrm{d}t} &\!\!=\!\!& -c_2 \rho_j^2 + K \left( -c_1 \rho_j^2 + \rho_j\rho_{j-1} g(\phi_j) \right) + \sigma \left( X_j \eta_j^Y - Y_j \eta_j^X \right)\\ \nonumber
\end{eqnarray}
where the auxiliary functions f and g have been introduced in equations~(\ref{functions_f_g}). The sum of two Gaussian variable is itself a Gaussian variable, whose average value is the sum of the two previous average value while its variance is the quadratic sum of the variances. Therefore we can introduce two new Gaussian delta correlated and zero mean white noise variables $\xi_j^{\rho}$ and  $\xi_j^{\theta}$ such that their standard deviations are
\begin{equation}
\Sigma_{\rho, \theta} = \sqrt{X_j^2 + Y_j^2} = \rho_j
\end{equation}
This leads to the final Langevin equations in polar form
\begin{eqnarray}
\frac{\mathrm{d}\rho_j}{\mathrm{d}t} &\!\!=\!\!& \rho_j - \rho_j^3 + K \left( -\rho_j + \rho_{j-1} f(\phi_j) \right)   + \sigma \xi^{\rho}_j\\
                                         \frac{\mathrm{d}\theta_j}{\mathrm{d}t} &\!\!=\!\!& -c_2  + K \left( -c_1  + \frac{\rho_{j-1}}{\rho_j} g(\phi_j) \right) + \frac{\sigma}{\rho_j}\xi^{\theta}_j \\ \nonumber
\end{eqnarray}
which display a multiplicative but delta correlated zero-average noisy term. In our power spectrum analysis, conducted expanding near the limit cycle solutions, this multiplicative component can be safely approximated by its limit cycle value, making the dominant noise component additive. 

\newpage
\newcommand{\newblock}{}

\end{document}